\documentclass[pra,twocolumn,aps,showpacs,flushbottom]{revtex4-1}
\usepackage{bm}
\usepackage{psfrag} 
\usepackage{graphicx}

\usepackage{amsmath}
\usepackage{amssymb}
\usepackage{amsthm} 
\usepackage{bm}
\usepackage{color}

\newcommand{\bea}{\begin{eqnarray}}
\newcommand{\ea}{\end{eqnarray}}
\newcommand{\eea}{\end{eqnarray}}

\newcommand{\sumint}[1]

\begin{document}

\title{Condensate fragmentation as a sensitive measure of the quantum many-body behavior of bosons with long-range interactions}

\author{Uwe R. Fischer,$^1$ Axel U.\,J. Lode,$^2$ and Budhaditya Chatterjee$^1$}

\affiliation{$^{1}$Seoul National University, Department of Physics and Astronomy \\  Center for Theoretical Physics, 151-747 Seoul, Korea}
\affiliation{$^{2}$Condensed Matter Theory and Quantum Computing Group, Departement f\"ur Physik \\ Universit\"at Basel, Klingelbergstr.\,82, 4056 Basel, Switzerland}

\begin{abstract}
The occupation of more than one single-particle state and hence the emergence of fragmentation is a  many-body phenomenon universal to systems of spatially confined interacting bosons. In the present study, we investigate the effect of the range of the interparticle interactions on the fragmentation degree of one- and two-dimensional systems. We solve the full many-body 
Schr\"odinger equation of the system using the recursive implementation of the multiconfigurational time-dependent Hartree for bosons method, R-MCTDHB. The dependence of the degree of fragmentation on dimensionality, particle number, areal or line density and interaction strength is assessed. It is found that for contact interactions, the fragmentation is essentially density independent in two dimensions. However, fragmentation increasingly depends  
on density the more long-ranged the interactions become. 
The degree of fragmentation is increasing,  keeping the particle number $N$ fixed,
when the density is decreasing as expected in one spatial dimension. We demonstrate that this remains, nontrivially, true also for long-range interactions in two spatial dimensions. We, finally, find that within our fully self-consistent approach, the fragmentation degree, to a good approximation, decreases universally as $N^{-1/2}$ when only $N$ is varied.

\end{abstract}


\maketitle

\section{Introduction}

The many-body behavior of systems of interacting bosons 
can in many cases be essentially characterized 
by a single ``diluteness parameter'' expressing the importance of interactions \cite{Pethick}. 
For most experimentally produced samples of ultracold bosonic gases, describing them by such a (small) diluteness parameter essentially yields the correct physics, because the mean interparticle distance is much smaller than the range of the interparticle interaction potential. 
Archetypical theoretical methods employing the diluteness of the gas  
are the Gross-Pitaevskii mean-field theory \cite{Pethick} and, with (quantum) fluctuations
incorporated to lowest order, Bogoliubov theory \cite{Bogoliubov}. These approaches assume that a single macroscopically occupied orbital forms the many-body ground state, and hence that a single creation operator
in the field operator expansion is sufficient. 

In particular in one and in three spatial dimensions (1D and 3D), 
diluteness criteria are well established for integrable as well as nonintegrable 
interparticle interactions with a potential $V_{\rm int} (r)\propto 1/r^n$, \cite{PitaSandro}. Integrability here means the (infrared) convergence of the integral
$\int V_{\rm int} (r) d^Dr$, implying $n>D$, where $D$ is the spatial dimension.  In the following, we use the terminology of interactions being short- and long-range interchangeably with them being integrable and nonintegrable, respectively. 
It is a widely accepted fact that for short-range interactions, the gas is more dilute and theories such as Bogoliubov mean-fields are applicable for low densities in 3D and, conversely, for large densities in  1D.
For long-range interactions, that is in particular for Coulomb interactions, the gas cannot be considered as dilute when the density {\em decreases} in 3D. A consequence of this fact is, for example, the formation of a Wigner crystal  \cite{Wigner}.  The 3D case can however for long-range interactions like Coulomb be described by Bogoliubov theory for sufficiently {\em high} densities \cite{Foldy}. On the other hand, in one dimension, high densities, in combination with Coulomb interactions, that is 
$n=1$,  lead to a Tonks-Girardeau gas \cite{Astra}. 
For strongly repulsive contact interactions in 1D the density profile can either be captured by a Thomas Fermi or a Tonks-Girardeau local equation of state for high or low densities, respectively \cite{Dunjko}.

{\em Two spatial dimensions} (2D), which are the main focus in what follows, represent a 
crossover case between the two limiting behaviors of 1D and 3D with respect to the effective diluteness of the gas,  
and therefore constitute a sensitive probe for many-body effects. In 2D, correlations can potentially play a significant role for both, integrable and nonintegrable interactions.  
For integrable interactions like contact pseudopotentials it is known that the diluteness parameter in 2D  is essentially independent on the areal density $\rho$, involving a double logarithm $\ln\ln[\rho a_s^2]$, where $a_s$ is the $s$-wave scattering length \cite{Fisher}. 
In the infinitely extended homogeneous gas, two dimensions represent the marginal case for the existence of condensates \cite{MerminWagner,Hohenberg,Fischer,Castin}; 
at any finite temperature condensation is absent, while at zero temperature condensation is possible albeit for extremely small densities \cite{Schick}. 
Two-dimensional Bose gases have been comprehensively studied for short-range 
interactions, cf. the review \cite{Anna}. However, 
comparatively little is known for long-range interactions, apart from the case 
where gauge fields of constant large flux are externally applied to two-dimensional electron gases, where the theoretical foundation of the fractional quantum Hall effect in the form of the eponymous Laughlin wavefunction  \cite{Laughlin} has spurred both tremendous experimental and theoretical activity \cite{Stoermer}.  
We also note in this regard that the many-body physics of dipolar bosons in 2D, cf., e.g., \cite{Baranov} 
does not qualify with the presently employed terminology as being genuine ``long-range,'' due to the integrability of the interaction potential of (polarized) dipoles in two spatial dimensions. 

The question how the diluteness of a bosonic many-body system depends on the power law tail of the interaction potential 
has thus not yet been addressed generally, in particular 
for the two-dimensional case. In the present work, we aim to fill this gap by employing the recursive implementation of the multiconfigurational time-dependent Hartree for bosons (R-MCTDHB) \cite{ultracold}, to solve the many-body problem 
with both integrable and nonintegrable interaction potentials of varying power law $n$.
 
The phenomenon of fragmentation, cf., e.g., \cite{Penrose,Leggett,Mueller,Alon08,Bader}, in which more than one field operator mode becomes macroscopically occupied, so that Bogoliubov mean-field theory necessarily breaks down, has previously in particular been extensively studied in one dimension. For instance, a stepwise increase of the number of fragments takes place in an asymmetric double well \cite{AlonPRL05}. Because of the strongly increased demand in computational resources the two-dimensional and three-dimensional cases have only recently attracted attention due to the availability of novel numerical methods, see for instance \cite{ultracold}. Fragmentation has been found in the case of an effective potential barrier \cite{Streltsov} induced by the long-range interactions or the dynamics above a condensed ground state \cite{Klaiman}. 

We focus below on studying fragmentation as an ``intrinsic'' property of the many-body system in a single (e.g. hard wall or harmonic) trap. In this case, fragmentation is termed intrinsic, because it is not assisted by the one-body potential in the Hamiltonian. This is in contrast to the cases where fragmentation is ``extrinsic'', i.e., assisted by the one-body potential as for instance in double wells or optical lattices. We refer to intrinsic fragmentation as a genuine many-body effect, because it originates solely in the interparticle interactions and thus does not essentially depend on the way in which the bosons are confined.

Exactly solvable cases for benchmarking the numerics are very scarce and significantly more specialized in spatial dimensions higher than one. 
For the one-dimensional case, exactly solvable cases are available such as the Tonks-Girardeau gas and the Lieb-Liniger and Calogero-Sutherland models, as reviewed in \cite{Cazalilla}.  
Analytical ground state solutions in two dimensions have been obtained for instance for the bosonic variant of Laughlin's wavefunction for particles in an (effective) magnetic field with hardcore interaction \cite{Paredes}, or the variant of the Calogero model considered in \cite{Khare}. In any spatial dimension, Richardson's pairing model employing a contact pairing
interaction potential allows for exact solutions \cite{Dukelsky}.
Numerically, it has been shown using the harmonic interaction model \cite{HIM1,HIM2} and its time-dependent generalization that the approach of (R-)MCTDHB can solve the generally time-dependent many-body problem 
to an in principle arbitrary degree of accuracy in 1D \cite{Axel_exact} and also in 2D \cite{Axel_book}.  
 
 In what follows, we exploit the fact that condensate fragmentation is a genuine many-body phenomenon whose very existence relies on both first- {\em and} higher-order correlations. 
We show that the fact of the gas being no longer dilute, and hence   
Bogoliubov theory breaking down, depends strongly on the power law of two-body interactions.
In the present study, a system will therefore be considered dilute if it is accurately described by a mean-field theory because fragmentation is absent. On the other hand, a system is said to be not dilute 
(or strongly correlated), if fragmentation is present and mean-field theories are not applicable.
We thus use the degree of fragmentation as an indicator of the ``many-bodyness'' of an interacting
quantum gas: When fragmentation becomes significant, the description of the system on an 
effective single-particle level, that is within mean-field theories such as the Bogoliubov theory, 
becomes inapplicable. 
 
We will show that, indeed, the degree of fragmentation is both small and essentially density-independent for contact interactions.
However, the more long-ranged the interactions become, the more significantly the fragmentation degree depends on the density
and exceeds the contact interaction value at sufficiently small densities. 
The degree of fragmentation is thus {\em increasing} when density of particles in the gas disk decreases. While this is a well-known fact in one dimension and for contact interactions, in two dimensions and for long-range interactions this result is a nontrivial and novel manifestation of many-body physics. 

In addition, we make explicit that besides the density the total number of particles $N$ is an 
independent parameter by which to assess the mean-field behavior of the system.
We demonstrate that in any dimension, and for any power law of interactions, the degree of fragmentation, and thus the degree to which a genuine quantum many-body nature of the interacting system is manifest, is reduced with increasing $N$.
This markedly differs from the functional behavior of the degree of fragmentation upon varying $\rho$, which strongly depends 
on spatial dimension and power law of interactions. 

\section{Model and Methods} 

\subsection{Setup} 

The $N$-body Hamiltonian reads  
\begin{equation}
\hat{H} = \sum_{i=1}^{N} \hat{h}(\bm{r}_i) + \sum_{i<j=1}^N \hat{V}_{\rm int} (\bm{r}_{i},\bm{r}_{j}).
\end{equation}
Here, $\hat{h}(\bm{r}_i)=\left( -\frac{1}{2} \nabla_{i}^{2} + V_{\rm ext}(\bm{r}_{i}) \right)$, is the one-body Hamiltonian that contains the kinetic and potential energy of the particle with index $i$ and  $\hat{V}_{\rm int}(\bm{r}_{i},\bm{r}_{j})$ is the two-body interaction for a pair of particles at $\bm{r}_i$ and $\bm{r}_j$.
For reasons of  computational convenience, we rescale the Hamiltonian, by using a typical length scale of the system $L$, by  
$\frac{\hbar^2}{2mL^2}$. This defines the units of length and energy we employ; all quantities in what follows are dimensionless.

We consider a two-dimensional hard-wall disk potential specifically given by  
\begin{equation}
V_{\rm ext}(r) = \left\{
   \begin{array}{ll}
   0 & r \leq a \\
   1000 & r > a
   \end{array}
\right. ,
\label{eq:hard_sphere_pot}
\end{equation}
The areal density for $N$ particles is thus given by $\rho = \frac{N}{\pi a^2}$.
Note that most recent advances in engineering atom traps allow for hard-walled ``box'' traps, enabling the 
experimental study of the ground state of homogeneous Bose gases, in various spatial dimensions 
\cite{Zoran,Corman}. 

In view of the convergence issues for a true $\delta$ potential in two dimensions \cite{doganov,friedman}, 
the contact interaction is modeled as a normalized Gaussian as follows 
\begin{equation} 
V_{\rm int}(R=|\bm{r}_i -\bm{r}_j|) = g\frac{\exp\left(-\frac{R^{2}}{2\sigma^{2}}\right)}{2 \pi \sigma^2}, \label{Vcontact}
\end{equation}
with width $\sigma = 0.25$ chosen, such that it reproduces the physics of a contact interaction \cite{doganov}. The long-ranged interaction potential of two particles at $\bm{r}_i$ and $\bm{r}_j$, respectively, is modeled by the regularized expression 
\begin{equation}
V_{\rm int}(R=|\bm{r}_i -\bm{r}_j|) = \frac{g}{R^n + \Delta^n}, \label{Vlongrange}
\end{equation}
where $\Delta (= 0.07)$ is the cutoff and $n$ is the power law of the long-range interaction potential.

Fragmentation is computed by calculating the natural orbital occupation numbers $N_i$, 
which are the eigenvalues of the one-body density matrix 
\begin{eqnarray}
 \rho^{(1)}(\bm{r}'\vert \bm{r}) 
  &=& \sum_\alpha N_\alpha \phi_\alpha^{(NO),*}(\bm{r}') \phi_\alpha^{(NO)}(\bm{r}), \label{RDM}
\end{eqnarray}
with the eigenvectors $\phi_\alpha^{(NO)}$ representing the natural orbitals. 
The degree of fragmentation (for two modes) is  then defined as \cite{Bader} 
\bea
{\mathcal F} = 1-\frac{|N_1-N_2|}N,
\ea 
where $N_{1,2}$ are the two largest natural orbital occupation numbers, defined by Eq.\,\eqref{RDM}.  
We note that we have verified that in the parameter ranges under investigation below, the occupation of a possible third orbital remains negligibly small, so that performing a truncation after two modes in the field operator expansion is justified.

\subsection{Numerical Methods}

We use the multiconfigurational time-dependent Hartree method for Bosons (MCTDHB) \cite{Alon08} in its recent recursive implementation (R-MCTDHB) \cite{ultracold} to compute the solution of the Schr\"odinger equation to an in principle arbitrarily large degree of accuracy. The MCTDHB method uses the following ansatz for the wave function 
\begin{equation}
 \vert \Psi \rangle = \sum_{\vec{n}} C_{\vec{n}} (t) \vert \vec{n}; t \rangle. \label{Ansatz}
\end{equation}
Here, $\vert \Psi \rangle$ is expanded as a linear combination of time-dependent, fully variationally determined and symmetrized many-body basis states, 
\begin{equation}
 \vert \vec{n} ; t \rangle = \frac{1}{\sqrt{\prod_{\alpha=1}^M n_\alpha !}} \prod_{\alpha=1}^M \left( b_\alpha^\dagger(t) \right)^{n_\alpha} \vert \textrm{vac} \rangle,
\end{equation}
where a vector notation was invoked for the occupations, $\vec{n}=(n_1,...,n_M)^T$, which fulfill $N=\sum_\alpha n_\alpha$. The many-body basis states $\vert \vec{n}; t \rangle$ are obtained by applying a symmetrization operator to Hartree products that are built from at most $M$ distinct orthonormal 
orbitals $\lbrace \phi_\alpha(\bm{r};t); \alpha=1,...,M \rbrace$. With this ansatz one tackles the time-dependent many-body 
Schr\"odinger equation, using the time-dependent variational principle \cite{Kramer}. As the result of the variation of the action functional, two sets of equations of motion, one non-linear integro-differential set for the time-evolution of the $M$ single-particle basis states or orbitals $\lbrace \phi_\alpha(\bm{r};t); \alpha=1,...,M \rbrace$ and one linear set for the time-evolution of the $\binom{N+M-1}{N}$ coefficients $\lbrace C_{\vec{n}}(t) \rbrace$, are obtained, see Ref.\,\cite{Alon08} for a detailed discussion. The two sets are coupled, because the time-evolution of the 
orbitals depends on the matrix elements $\rho_{\alpha\beta}(t),\rho_{\alpha\beta\gamma\delta}(t)$ which are functions of the coefficients, whereas the time-evolution of the coefficients depends on the matrix elements of the one-body and two-body Hamiltonian, $h_{\alpha\beta}, W_{\alpha\beta\gamma\delta}$. The simultaneous and self-consistent solution of the system of MCTDHB equations of motion for both $\lbrace C_{\vec{n}}(t) \rbrace$ and $\lbrace \phi_\alpha(\vec{r};t) \rbrace$ is equivalent to solving the full (time-dependent) many-boson problem \textit{if} convergence with respect to the number of variational parameters is achieved, see \cite{Axel_exact,Axel_book}. This is a consequence of the formal exactness of the ansatz in Eq.\,\eqref{Ansatz}; in the case of $M \rightarrow \infty$, it covers the full $N$-boson Hilbert space. Note that in the case of $M=1$, Equation \eqref{Ansatz} is a mean-field product state and therefore the set of MCTDHB equations boils down to the time-dependent Gross-Pitaevskii equation. 

The solution of the coupled set of equations for $M>1$ ß 
is a numerically demanding task for which the R-MCTDHB package \cite{ultracold} is used.

\section{Integrable and nonintegrable Interactions}
\subsection{Two-dimensional disk geometry} 
\subsubsection{Fragmentation for various power laws of interaction}

We first remark, that for the present discussion, and especially for the presently considered mesoscopic systems, $N$ and $\rho$ have distinct influence on the many-body state of the system. In the first-quantized form, $N$ is a fundamental parameter of the many-body Hamiltonian while  $\rho$ is the ratio $\rho = N/A$; here, the area $A$ depends on the external potential. A change in $N$ affects both the external potential and the interaction term while changing $A$ requires changing only the  potential term and is independent of the interaction term. Since the $N$- and $A$-dependence of the system are different, the value of $\rho$ does not uniquely determine the many-body wave function, which also explicitly depends on the value of $N$.

Hence, to analyze the density dependence of fragmentation independent from the influence of $N$, we keep $N=100$ fixed and first vary the area $A$ of the disk, by changing its radius $a$,  in the 2D disk potential Eq.\,\eqref{eq:hard_sphere_pot}. 

The density dependence of fragmentation is assessed using a contact interaction and interactions with power laws $n=1,2,3$ in Eq.\,\eqref{Vlongrange}. In principle, a comparison between different interaction potentials can be made  with respect  to the same coupling constant $g$. However, a more useful comparison can be  obtained by using the following integrated measure of the 2D interaction potential $V_{\rm eff} \equiv 2\pi \int V_{\rm int}(R)R dR$ ($\propto g$), since $V_{\rm int}$ enters the many-body Hamiltonian as the kernel of the integrals $W_{\alpha\beta\gamma\delta}$, see Ref.\,\cite{Alon08} and Eq.\,\eqref{wksql} below. The coupling constants are then renormalized such that they correspond to the same $V_{\rm eff}$ for different interaction potentials. 

\begin{figure}[t]
\includegraphics[width=0.95\columnwidth,keepaspectratio]{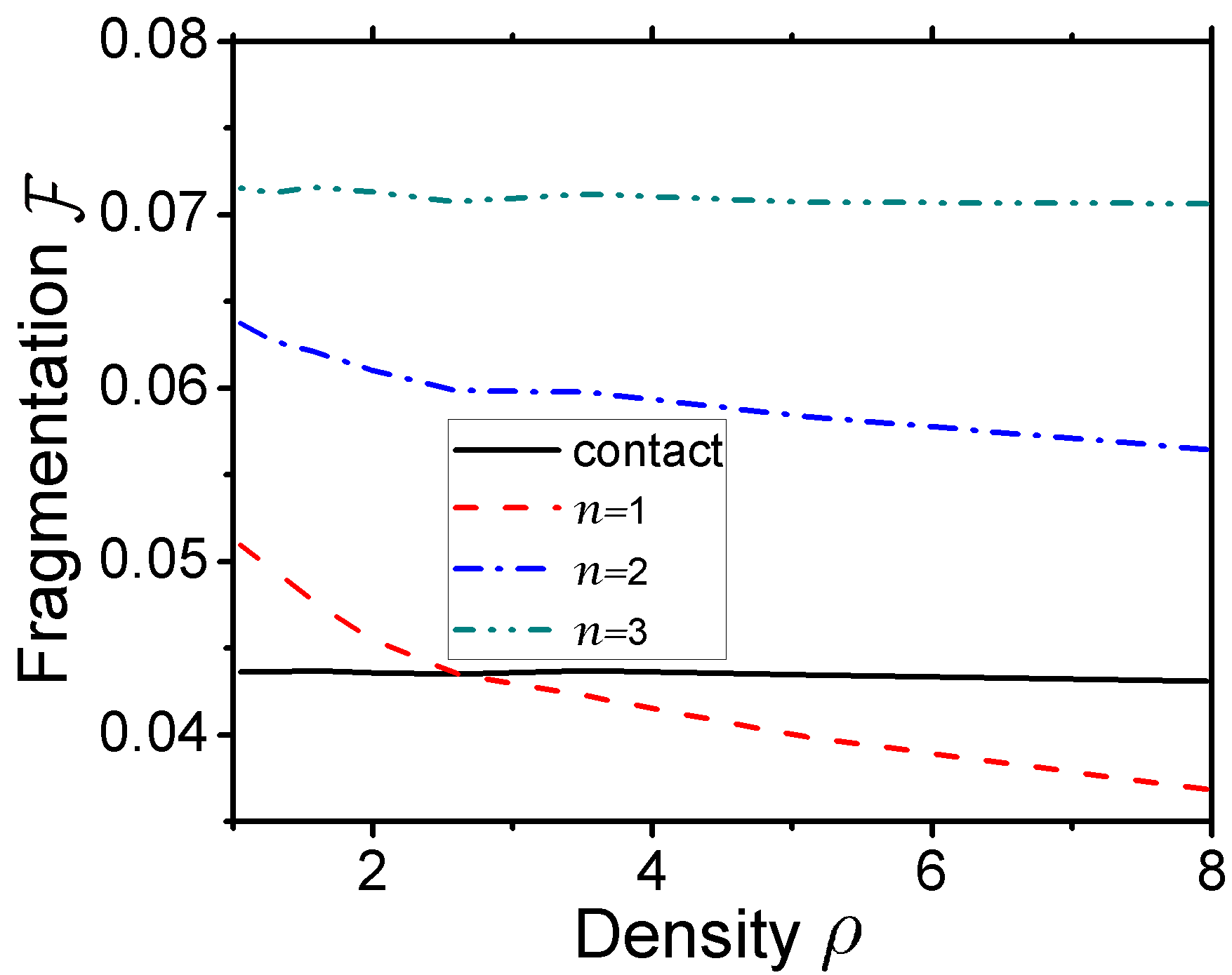}
\caption{(color online): The variation of fragmentation with 2D density $\rho = \frac{N}{\pi a^2}$ for 
$N=100$ particles, comparing contact interaction and interactions of power law $n=1,2,3$, at fixed $V_{\rm eff}=2\pi\times 11.6$. 
\label{cap:F_compare_N100g11_64renorm}}
\end{figure}

In Fig.\,\ref{cap:F_compare_N100g11_64renorm}, we compare the variation of fragmentation $\mathcal F$ with density $\rho$, holding $V_{\rm eff}=2\pi\times 11.6$ constant.  For contact interaction (cf. Eq.\,\eqref{Vcontact}), there is hardly a clearly discernible variation of $\mathcal F$ with $\rho$.  For long-ranged interactions ($n=1,2$ in Eq.\,\eqref{Vlongrange}), on the other hand, the degree of fragmentation decreases with $\rho$. For the 2D-integrable $n=3$, and $n>3$ (not shown),  the influence of density on fragmentation is again minimal. 

This leads us to conclude that the influence of density on the degree of fragmentation for two spatial dimensions depends on the integrability of the interaction potential; for integrable interaction Eqs.\,\eqref{Vcontact}, and \eqref{Vlongrange} for $n\geq3$, the variation of $\mathcal F$ with $\rho$ is minimal, while for nonintegrable interactions, $n=1,2$ in Eq.\,\eqref{Vlongrange}, the fragmentation degree decreases with density. 
We also note that for $n=1$, the most long-ranged case, the extent of the decrease of $\mathcal{F}(\rho)$ is significantly stronger than for $n=2$, that is the more long ranged the interaction is, the stronger
$\mathcal F$ varies with density.  
The $n=1$ curve, while starting from higher values for small densities, crosses the curve for contact interactions, decreasing to a lower value of $\mathcal F$ for higher densities. The $n=2$ curve again starts with a higher value of $\mathcal F$ for low density, decreasing slowly and approaches the contact interaction curve for large densities. For $n=3$, 
$\mathcal{F}(\rho)$ is highest for all densities.
Finally, for sufficiently large densities, the $n=1$ interaction is the least fragmented and the integrable $n=3$ interaction is the most fragmented.
We have verified that these qualitative features are similar for any number of particles $N$ and interaction coupling $g$ with only the value of fragmentation $\mathcal F$ being different.

\subsubsection{Effective Thomas-Fermi parameters}

The degree of fragmentation for a fixed $N$ depends on the interplay between the kinetic energy and interaction energy. 
Thus, an important measure in determining the fragmentation of the system is the Thomas-Fermi parameter represented by the interaction energy and kinetic energy, $P_{\rm TF}
=\frac{E_{\rm int}}{E_{\rm kin}}$. 

\begin{figure}[t]
\includegraphics[width=0.9\columnwidth,keepaspectratio]{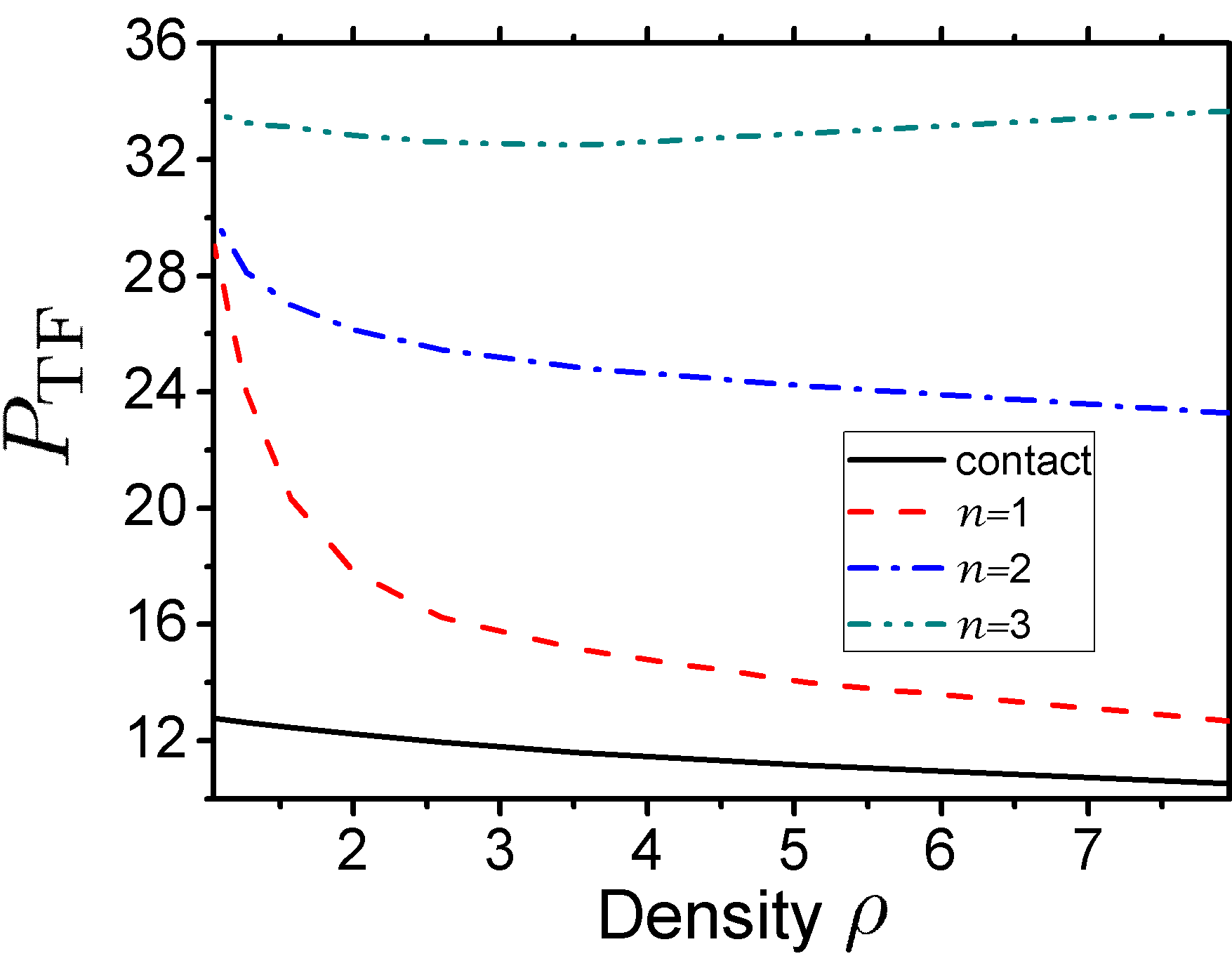}
\caption{(color online): The variation of the Thomas-Fermi parameter $P_{\rm TF}(\rho)$ with 2D 
density $\rho$ for a fixed $N=100$, for contact interaction and long-range interactions of degree $n=1,2,3$, in two spatial dimensions.  
\label{cap:energy}}
\end{figure}

Figure \ref{cap:energy} shows the variation of the Thomas-Fermi parameter $P_{\rm TF}$ with density $\rho$ for fixed $N=100$. One can observe a significant difference between integrable and nonintegrable interactions in particular in the low density regime. For contact interaction, there is a very slow decrease of $P_{\rm TF}$ with $\rho$ throughout the entire density regime; $n=3$ shows similar behavior, with $P_{\rm TF}$ remaining roughly constant with $\rho$. A strikingly different functional dependence is seen for $n=1$. While the  large $\rho$ behavior is similar to that of contact interactions, for small $\rho$, there is a very sharp increase of $P_{\rm TF}$ with decreasing density. A similar tendency (albeit to a  smaller degree) is observed for $n=2$. This strong increase of the $P_{\rm TF}$ for decreasing density, implying the stronger influence of the interaction energy as compared to the kinetic energy, manifests itself in the increasing degree of fragmentation with lowering density seen for nonintegrable 
interactions, cf.\,Fig.\,\ref{cap:F_compare_N100g11_64renorm}.

To obtain a qualitative understanding of the above discussed dependence on $\rho$, we note that in the mean-field regime for contact interactions, both the interaction and kinetic energy go linearly with density and  $P_{\rm TF}$ does not vary with density. In a many-body context, the equally linear dependence of kinetic and interaction energy does not strictly hold and deviations are observed.  In our case, the interaction energy goes slightly faster than the kinetic energy leading to  the small gradual decrease of $P_{\rm TF}$ with density. The difference from the mean-field behavior is, however, much more significant for $n=1$ and $n=2$, with the largest effect for $n=1$. Here, especially for $n=1$, the variation of $E_{\rm int}$ with $\rho$ is not of a simple power law form and overall increases slower compared to $E_{\rm kin}$, leading to the rapid rise of $P_{\rm TF}$ at small densities. For example, when $n=1$, $E_{\rm int} \sim \rho^{0.45}$ while 
$E_{\rm kin} \sim \rho^{0.7}$. Similarly, for $n=2$,  $E_{\rm int}\sim \rho^{0.8}$ while $E_{\rm kin}\sim \rho^{0.9}$.

\subsubsection{Density profiles}
\label{densitypattern}

\begin{figure}[t]
\includegraphics[width=1.025\columnwidth,keepaspectratio]{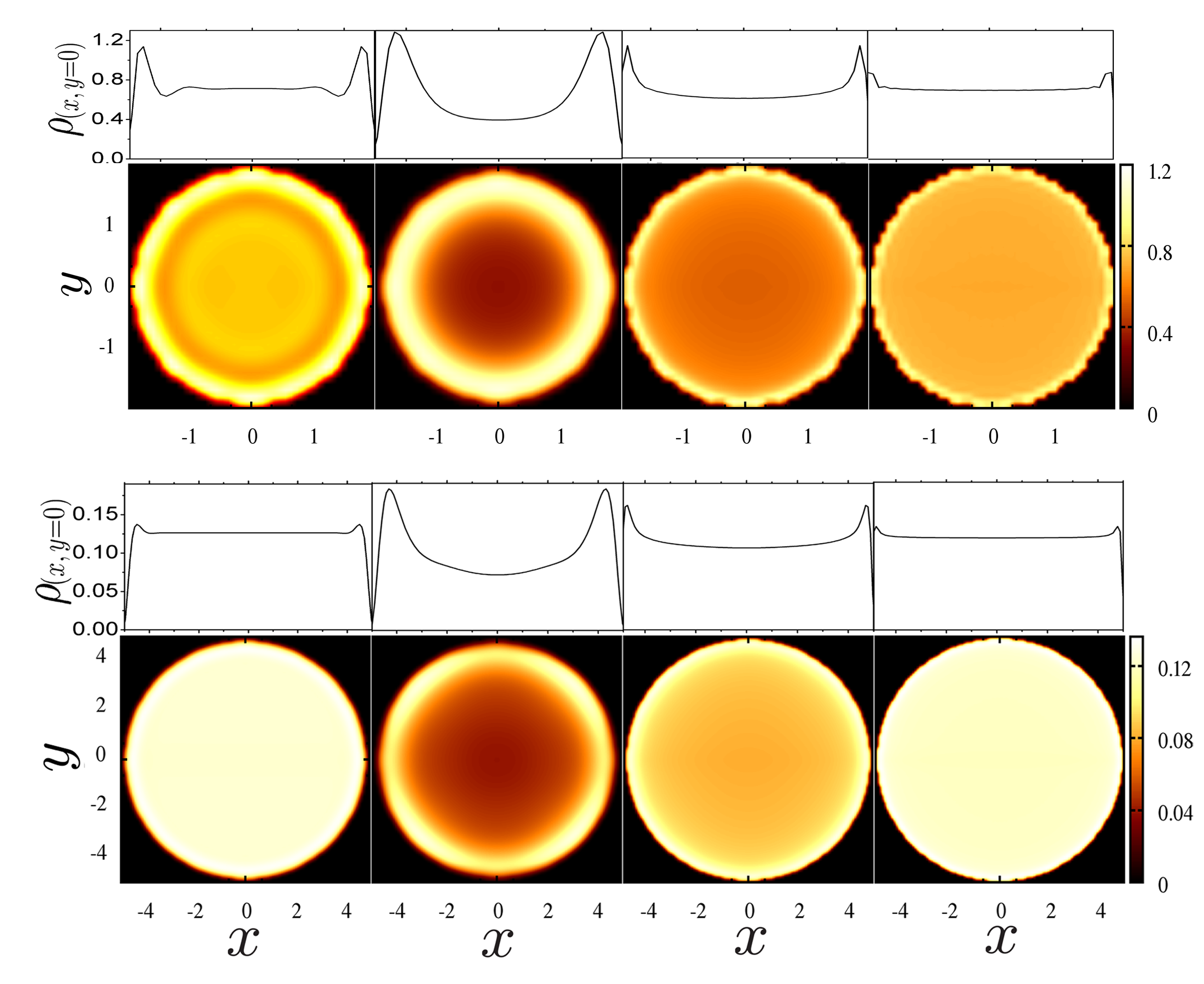}
\caption{(color online): 2D density distributions of the disk and  the respective 1D cut at $y=0$  for  \textit{Top:} average density $\rho=7.95$ ($a=2$), for contact interactions, $n=1$, $n=2$, $n=3$, from left to right.  \textit{Bottom:} average density $\rho=1.27$ ($a=5$), at the same values of $n$. Darker  and brighter color indicates  lower and higher density, respectively; cuts at y=0 are also shown. 
\label{cap:density}}
\end{figure}

The difference between the influence of the integrable and nonintegrable interactions on fragmentation 
can be connected to the spatial density profile. For large densities (small radius), the  influence of the repulsive interactions for all values of $n$ is felt on the entire disk, leading to a higher density in the rim of the disk (see Fig.\,\ref{cap:density} top panel). As the radius increases (thus lowering $\rho$), the influence of the  short-ranged interactions (contact, $n=3$) on the whole area of the disk is reduced, because the radius becomes much larger than any length scale associated to interactions, and the density hence is rendered homogeneous. The nonintegrable interactions ($n=1,2$), on the other hand, retain the long-range influence of the interaction for any radius of the disk. As a result, the equalization of the densities does not occur, and  the rim of the disk continues to have higher densities, as clearly visible in Fig.\,\ref{cap:density} (bottom panel).

This difference manifests itself in the occupation of the energy levels and thus the degree of fragmentation observed. The short range of integrable interactions leads to an essentially localized behavior and  long-range correlations between the bosons are very weak. Nonintegrable interactions, on the other hand, imply longer-ranged correlations: The sphere of the influence of interactions involves the contribution of a much larger number of bosons.  A given boson thus interacts with  all the bosons in the disk irrespective of the  latter's radius and hence feels  a stronger effective interaction. 
In more formal terms, the kernel of the matrix elements of the two-body interaction,  
\begin{equation}
 W_{\alpha\beta\gamma\delta}= \int d^2rd^2r' \phi_\alpha(\bm{r}) \phi_\beta(\bm{r'}) V_{\rm int} (\bm{r}-\bm{r'}) \phi_\gamma(\bm{r'}) \phi_\delta(\bm{r}) \label{wksql}
\end{equation}
has a larger support for smaller values of $n$ (longer ranges of interaction). 
Therefore the integrals $W_{ksql}$ are effectively larger such that the system tends to minimize the energy by occupying several $\phi_j$ and hence fragments.

\subsection{Comparison with the one-dimensional case} 
While in 2D, for contact interactions, we observe the absence of a change in fragmentation with density, in 1D there is a significant influence of the density, in line with what is known from exact diagonalization studies \cite{Cazalilla}.    
We compare a system of $N=100$ bosons in the disk geometry for the 2D case and $N=100$ boson in a 1D ``tube'' with periodic boundary conditions.

\subsubsection{Long-range interactions}
\begin{figure}[t]
\begin{center}
\includegraphics[width=0.95\columnwidth,keepaspectratio]{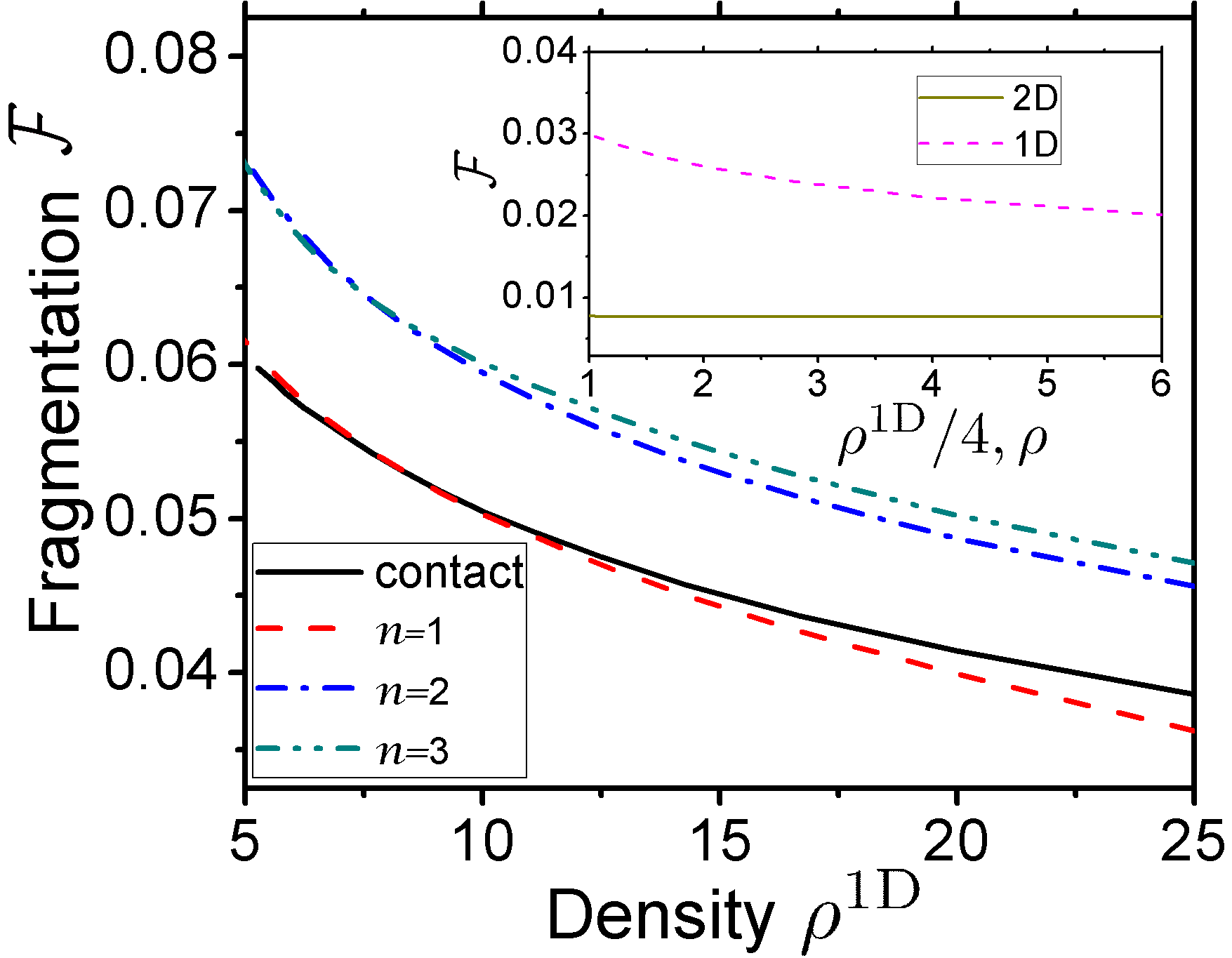}
\end{center}
\caption{(color online): Variation of fragmentation with density for $N=100$ 1D bosons with contact interactions and long-range interactions. Inset: Density dependence of fragmentation on dimensionality at equal density for $N=100$ 1D and 2D bosons, interacting with unit strength $g=1$. 
The 1D density $\rho^{\rm 1D}$ is scaled by $\frac{1}{4}$ to directly compare with the 2D range of densities $\rho$.
\label{cap:1d_renorm_compareN100g5.15_1d_2d}}
\end{figure}

Our simulations clearly reveal the difference between systems with long-range interactions in 1D and 2D. In Fig.\,\ref{cap:1d_renorm_compareN100g5.15_1d_2d} (main), we show the effect of the density $\rho^{\rm 1D}=\frac NL$ on fragmentation $\mathcal F$ in 1D for a constant  $V_{\rm eff}\equiv \int V_{\rm int}(R)dR =5.15$  
with $n$.  
Fragmentation decreases with density for all $n$. 
For $n=1$ in Eq.\,\eqref{Vlongrange}, the decrease with density is most rapid. For $n=2,3$ (which are integrable in 1D), the decrease is similar especially for larger densities. 
This is in complete accordance with our observations in the 2D bosonic gas: 
Concerning the dependence of $\mathcal{F}$ on $\rho$, integrable interactions demonstrate in either dimension similar behavior when compared to nonintegrable interactions.

\subsubsection{Contact interactions} 

For contact interactions, in Fig.\,\ref{cap:1d_renorm_compareN100g5.15_1d_2d} (inset), we compare the variation of fragmentation with density for 1D and 2D gases for $N=100$ and $g=1$ (for the sake of a direct comparison with the 2D computations, we rescaled the 1D density $\rho^{\rm 1D}$ in the inset of Fig.\,\ref{cap:1d_renorm_compareN100g5.15_1d_2d} 
by a factor $\frac14$).
We observe a substantial decrease of $\mathcal F$  for the 1D case with increasing density $\rho^{\rm 1D}$.  The comparison to 2D with its absence of a significant variation of $\mathcal F$ with density again highlights the marginal nature of the two dimensional system.  We note that in 3D, we anticipate that fragmentation is on one hand very small \cite{BaderPRA}, and on the other hand also decreases with decreasing density.

\section{Number dependence of fragmentation}
\begin{figure}[t]
\begin{center}
\includegraphics[width=0.95\columnwidth,keepaspectratio]{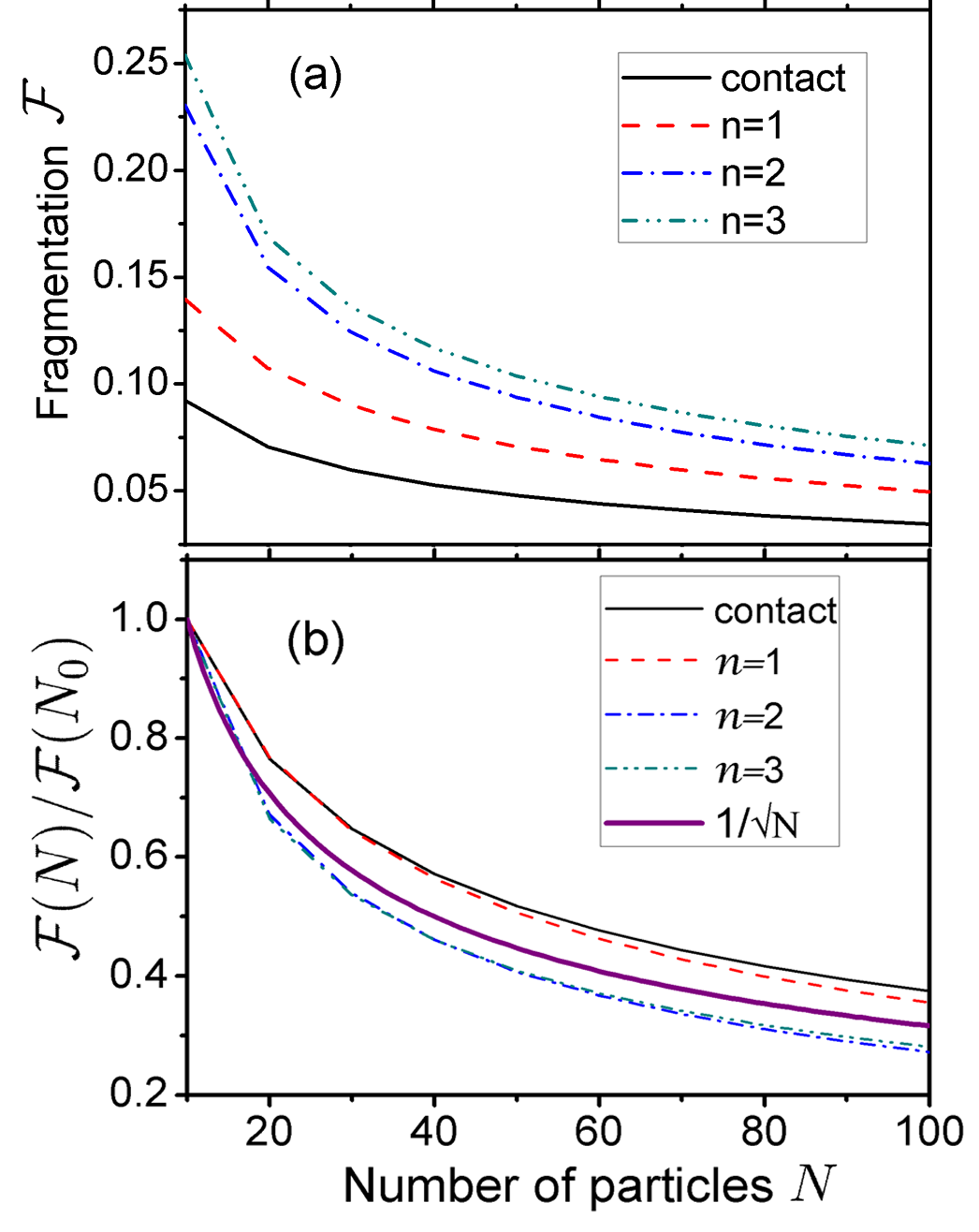}
\end{center}
\caption{(color online): (a) Number dependence of fragmentation for contact and long-range 
interactions for 2D. (b) Number dependence with fragmentation normalized to its value at $N_0=10$.
\label{cap:F_2d_N_var}}
\end{figure}

Fragmentation is a mesoscopic phenomenon in the sense that for fixed interaction coupling, increasing $N$ always leads to a decrease of fragmentation. We have observed that this is a universal phenomenon, and true for both integrable and nonintegrable interactions, for the presently considered ground states  of  1D and 2D bosonic gases contained in a single trap.

We illustrate in  Fig.\,\ref{cap:F_2d_N_var}(a) how the fragmentation $\mathcal F$ varies in 2D with the number of particles $N$ for contact and long-range interactions $n=1,2,3$, cf. Eqs.\,\eqref{Vcontact},\eqref{Vlongrange} at a fixed $V_{\rm eff}=2\pi\times 11.6$. A decrease in fragmentation is observed for all cases. 
Note that although with increasing $N$ and constant coupling $g$,  the mean-field interaction $V_{\rm MF}=(N-1)g$ increases, this does not lead to an increase of fragmentation $\mathcal F$. Instead, we find that the particle number $N$ dominates over
the effects of all other parameters affecting the fragmentation $\mathcal F$. That is,  the variation of $\mathcal F$ with other parameters like interaction strength $g$ and long-rangedness $n$ is subdominant in comparison.  
We stress that this qualitative behavior of the number dependence is universal in that it is true regardless of densities, interaction strength, and other parameters.
Note that this fact is not captured by a (single-parameter) variational theory \cite{FischerVar}, which demonstrates the sensitivity of fragmentation on the solutions of the  many-body equations being fully self-consistent.

In order to assess the relative decrease of fragmentation $\mathcal F$, we plot in Fig.\,\ref{cap:F_2d_N_var}(b) the $N$-dependence of $\mathcal F$ relative to the reference value set at  $N=10$. The contact interaction curve shows the smallest decrease with $N$, closely followed by $n=1$. The $n=2$ and $n=3$ show very similar dependence, and decay more rapidly with $N$ when compared to $n=1$.
For constant interaction coupling, the decrease of the fragmentation degree is
well approximated by $\mathcal F(N)/{\mathcal F}(N_0) \propto \frac{1}{\sqrt{N}}$, cf. Fig.\,\ref{cap:F_2d_N_var}(b). 
We finally mention that, when the effective mean-field interaction $V_{\rm MF}=(N-1)g$ were kept constant, then 
$\mathcal F$ falls off much faster (not shown), approximately as $\frac{1}{N}$.

\section{Conclusion} 

Using condensate fragmentation as a diagnostic tool, we have assessed the degree to which 
an interacting system of bosons can be considered to be dilute, and therefore describable by mean-field
theory and a single macroscopically occupied orbital. 
We have confirmed that the essential independence of the diluteness parameter in a 2D gas \cite{Fisher}
is reflected by an identical behavior of the degree of fragmentation $\mathcal F$  in a bosonic quantum gas, for integrable interactions. However, for nonintegrable interactions, a significant dependence of $\mathcal F$ on the density, namely a significant decrease of the degree of fragmentation  with increasing density obtains, and increasingly so for longer-ranged interactions.  
Due to the fact that fragmentation is a genuine many-body phenomenon, the degree of fragmentation thus represents a quantitative measure to which 
extent long-range interactions lead to strong correlations in two spatial dimensions. 
We have also found that in the large $N$ limit, in a self-consistent approach, mean-field theory again 
becomes valid. 
Fragmentation is therefore both a mesoscopic and intrinsically many-body phenomenon. 
Finally, the detection of the fragmentation we predict can be performed by measuring density-density correlations
after time of flight expansion of the cloud  \cite{Kang}.

\acknowledgments
The research of URF and BC was supported by the BK21 Program and the NRF of Korea, grant No. 2014R1A2A2A01006535. AUJL acknowledges financial support by the Swiss SNF and the
NCCR Quantum Science and Technology. 
 
\end{document}